# Modeling Composites of Multi-Walled Carbon Nanotubes in Polycarbonate

Prashant Jindal[1*], Meenakshi Goyal[2], Navin Kumar[3],

[1] University Institute of Engineering & Technology, Panjab University, Chandigarh-160014, INDIA

[2] University Institute of Chemical Engineering & Technology, Panjab University, Chandigarh-160014, INDIA

[3] Indian Institute of Technology, Roopnagar, Punjab, INDIA

**Abstract**

High strain rate experiments performed on multi-walled carbon nanotubes - polycarbonate composites (MWCNT-PC) have exhibited enhanced impact resistance under a dynamic strain rate of nearly 2500/s with composition of only 0.5 to 2.0% Multi walled carbon nanotubes(MWCNTs) in pure polycarbonate(PC). Similarly, hardness and elastic modulus under static loads resulted in significant increase depending upon the composition of MWCNTs in PC. The present work aims to analyze these results by correlating the data to fit expressions in generalizing the behavior of MWCNTs composition for MWCNT-PC composites under both static and impact loads. As a result we found that an optimum composition of 2.1 weight % of MWCNTs exhibits maximum stress resistance within elastic range under strain rates of nearly 2500/s for MWCNT-PC composites.

The results are critically dependent on the composition of MWCNTs. Theses results significantly deteriorate below and above a threshold composition. Further, a simple model based on Lennard –Jones 6-12 atom-atom based potential is formulated and used to compute static properties of pure as well as composite PC.

- Corresponding author, e-mail: jindalp@pu.ac.in

1. Introduction

Light weight and impact resistant materials are being systematically and extensively explored for determining their capability to sustain high and variable loading conditions to evaluate their usage in shock and impact absorption etc. [1, 2]. Making composites of low cost, easily available and machine-able base materials to composite with high quality materials of appropriate proportions has also been explored much more vigorously with good success in recent years.

Among the thermoplastic group of polymers, polycarbonates (PC) have attracted a great deal of attention due to their ability to be easily worked upon and mould ability. Their capability to resist temperature and impact makes them a common application material in house wares, laboratories and industries. A modification in their properties to suit specific requirements is an interesting proposition.

Ever since the synthesis of carbon nanotubes (CNTs) [3] and study that followed exploring mechanical and structural properties of CNTs [4-8], there has been wide ranging interests in scientific and engineering communities to exploit these for varying applications. The unusual mechanical strength of the carbon nanotubes revealing them as about 100 times stronger than steel motivates to fabricate and modify useful materials which are cheaply available in bulk form by embedding in these carbon nanotubes in various forms to make composites which have desired mechanical properties.

Keeping this in view, an investigation had been recently carried out by us [9] using Split Hokinson Pressure Bar to determine impact characteristics of MWCNT-PC composites. This work was followed up to measure their characteristics under static conditions by measuring the hardness and elastic constants of increasing MWCNT component in PC-MWCNT composites using nano-indenter[10]. As the results of these experiments were found to be highly motivating to assess the significance of the composites, we considered it worthwhile to suggest a simple basic model, especially for static properties in the first instance. Further, the significance of presenting the measured data of dynamic impact in terms of expressions and equations which represent this data was also considered important enough.

Molecular Dynamics(MD) computations have been widely used to for mechanical charazterization.. Molecular dynamics (MD) simulations have always been considered to provide reliable results in characterizing nanocomposites[11-13], however these computations do not provide enough insight and reasoning into what causes these characteristics. MD is limited to nanoscale and is costly in terms of computational inputs, which has promoted the development and usage of alternate approaches for characterizing CNT reinforced composites at microscale. One of these alternative approaches pertains to use of finite element method[14-18] which is used to validate the data on CNT reinforced composites.

Although there is some work in literature [19,20] that has modeled some features concerning nanomodification of polymeric resins for structural applications of obtaining high toughness even at low nanofiller volume fractions, yet such models are reasonably complicated. They claim that among these, nanoparticle debonding could take an important role either as a mechanism itself or as a trigger for phenomena like plastic void growth or matrix shear yielding. Therefore, the role of CNTs in PC can lead to toughening behavior not only because of energy dissipation through damage mechanisms at nanoscale, but also due to the toughened properties of CNTs and their networking with PC due to long lengths of CNTs. Recently, influence of nanoparticle size and shape on the mechanical properties of polymer nanocomposites has been well discussed by molecular dynamics simulations, including influence of nanoparticle size, loading, and shape on the mechanical properties of polymer nanocomposites. They report that spherical nanoparticles, whose size is comparable to that of a molecule of polymer, are more effective at toughening the polymer nano-composite than larger spherical particles. When comparing particles of spherical, triangular, and rod-like geometries, the rod-like nanoparticles emerge as the best toughening agents.

On the other hand, an organic solid like polycarbonate is a weak material which can be approximated to be governed by weak Van-der-Waals interactions between atoms of different molecules forming the polycarbonate. These atomic level interactions lead to the formation of molecule-molecule interactions and thus information of the solid structure and related properties can be obtained from minimum energy configurations. This approach has been quite successfully used to model static and bulk lattice properties of various organic solids [21-25]. The potential parameters that govern these atomic interactions have been obtained from their reproducibility of observed properties like lattice constant, bulk modulus and melting temperature etc. of some standard organic solids whose static and structural properties are well known. Therefore, these parameters are not specific to the material under consideration. These parameters for various atom-atom interactions are widely available in literature[22]. We use these without any modification to calculate interaction energy of pure and appropriate fractions of different MWCNTs compositions. Based on this, we further use the minimized energy configuration for calculation of density and Young's modulus for pure samples and composites.

## 2. Analyzing Dynamical Stress-Strain Data

Using least square fitting procedures, we have been able to make excellent representation similar to the experimental data for high strain rates. We find that cubic polynomials adequately fit major part of interesting range of stress-strain curves over a wide variation of concentrations of MWCNT in PC. The fitted results are presented in Fig. 1 and the best fitted parameters have been tabulated in Table 1.

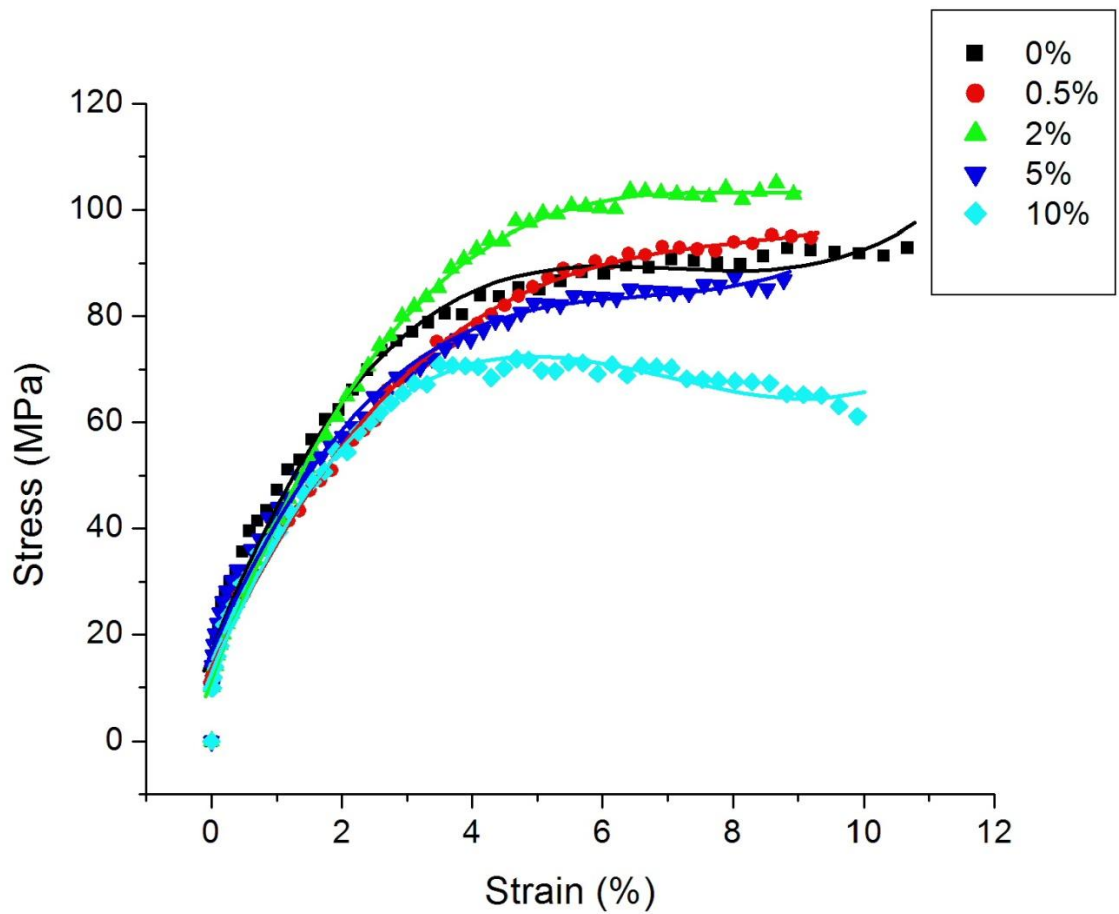

Fig. 1. Polynomial fits to the experimental data of stress as a function of strain for PC-MWCNT composites under high strain rate loading. The fitted parameters are given in Table 1.

For any given % composition of MWCNT in PC, the data for stress (p in Mpa) has been fitted to strain (x as %) as

$$p = A + Bx + Cx^2 + Dx^3 \qquad (1)$$

Similarly, to have an estimate of Young's modulus, we also plot in Fig. 2. Y, obtained from $p' = \frac{\partial p}{\partial x}$ as obtained from Eq. 1 as a function of strain. As the strain is in %, the Young's modulus Y will be 10 times $p'$ when expressed in GPa and is shown as

$$Y = p'/10 \text{ in GPa} \qquad (2)$$

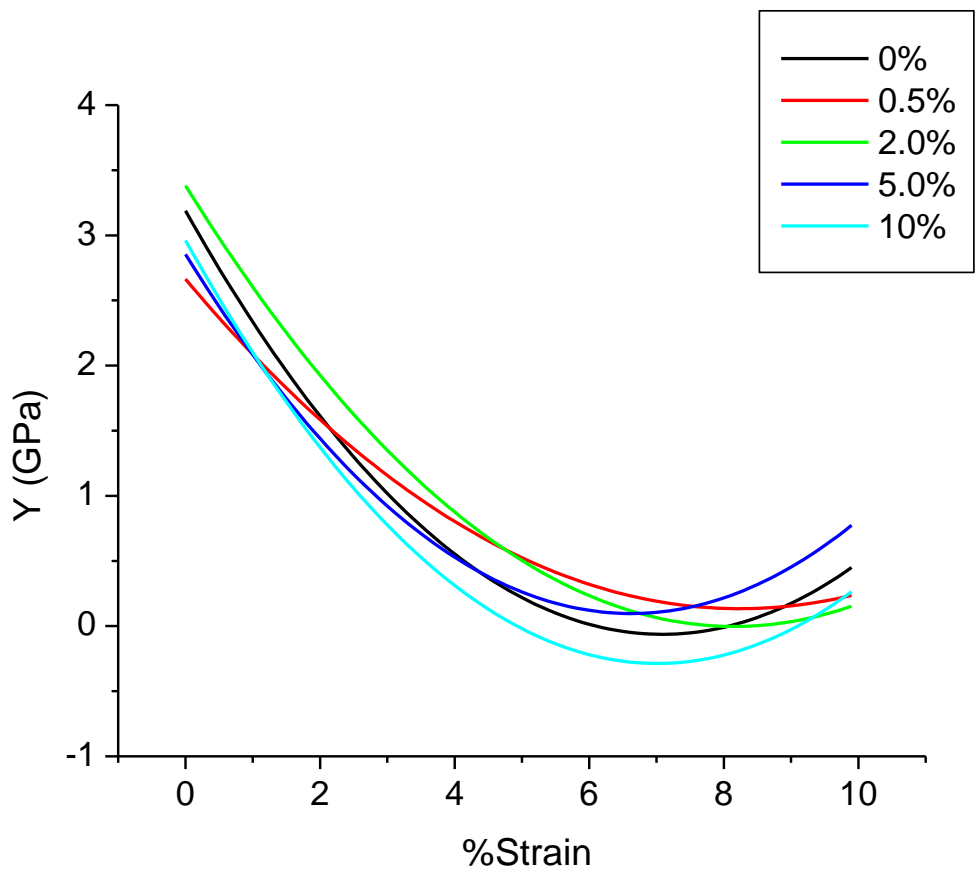

Fig. 2. Young's Modulus as a function of strain at a strain rate nearly 2500/s for various compositions of the MWCNT-PC composite Young's modulus at different strains and compositions from the dynamical impact data.

Table 1. Coefficients $A, B, C, D$ (Eq.1) for various concentrations of MWCNTs

| A | B | C | D | % MWCNT in PC |
|---|---|---|---|---|
| 16.70894 | 31.88236 | -4.58813 | 0.21582 | 0.0 |
| 13.56944 | 26.64232 | -3.07144 | 0.12425 | 0.5 |
| 11.44806 | 33.82367 | -4.15171 | 0.16973 | 2.0 |
| 16.55996 | 28.548 | -4.16623 | 0.20976 | 5.0 |
| 12.70309 | 29.61159 | -4.63311 | 0.22027 | 10.0 |

As one can easily observe from Fig.1, the cubic polynomial represents the dynamic impact data for strain rate 2500/s adequately well and it can be easily used with the parameters given in Table 1 to reproduce the experimental data.

We also plot Fig. 3, where the variation of maximum elastic stress with MWCNTs composition in PC is depicted under strain rates of nearly 2500/s. We notice from the Gaussian fitting that maximum elastic limit stress is obtained when the concentration of MWCNT is around 2.12%. The data fits well to the Gaussian

$$y = y_0 + \frac{A}{w\sqrt{\pi/2}} e^{-\frac{2(x-x_0)^2}{2w^2}} \tag{3}$$

Where $y_0$=67.26466, $A$ =109.54829, $x_0$=2.12425 and $w$ =3.62787.

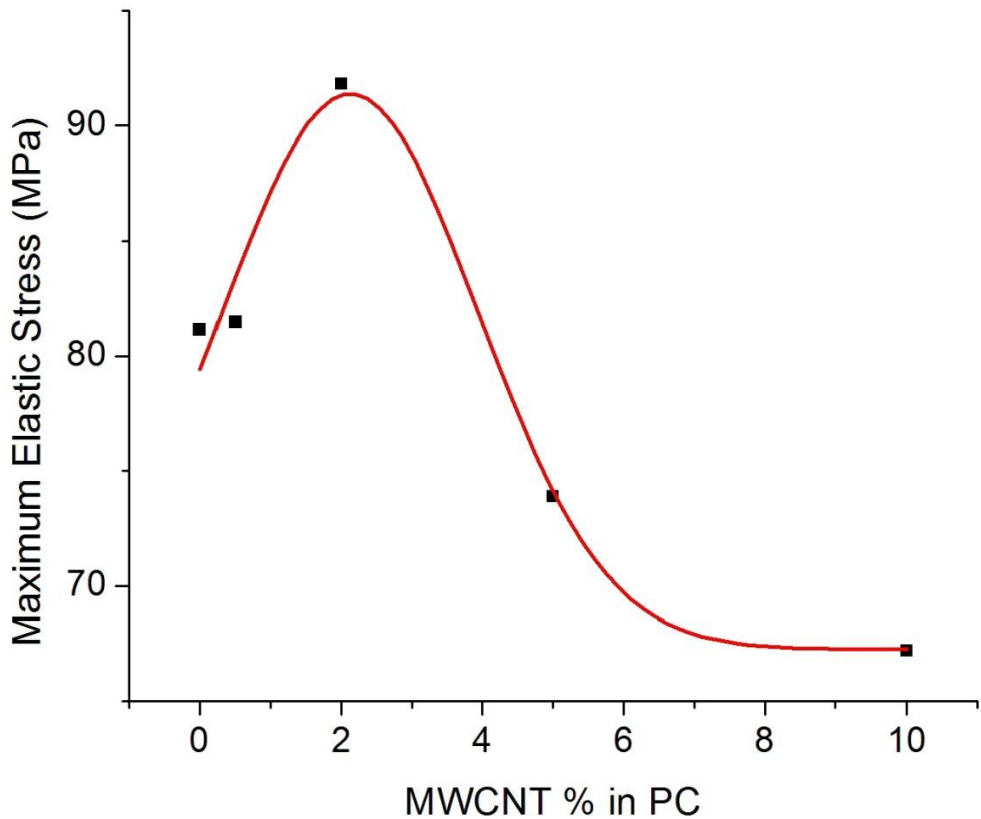

Fig. 3. Maximum elastic stress under dynamic impact of strain rate if about 2500/s as a function of MWCNT composition. The continuous line is a Gaussian fit.

### 3. Analyzing static stress data

We have explained about nano-indention tests performed on MWCNT-PC composites elsewhere[10] to evaluate mechanical properties-elastic modulus and hardness. Both these parameters are now fitted to polynomial expression which has been given in Fig. 4 and the data of fit in Table 2.

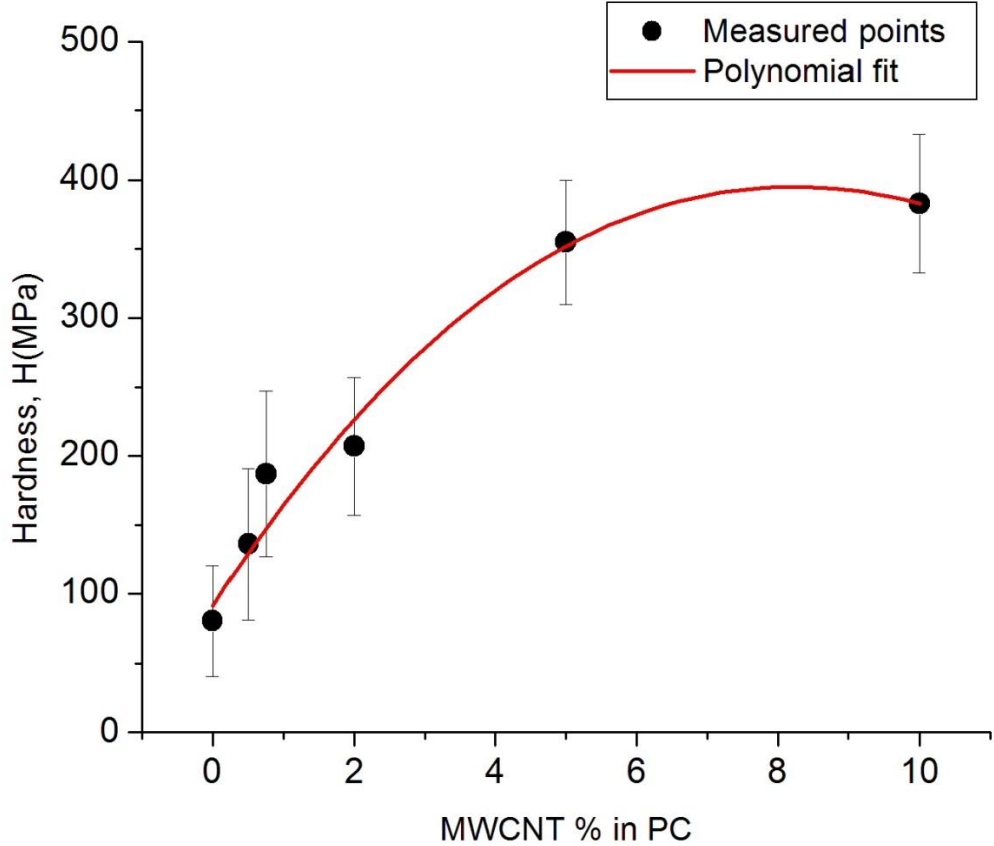

Fig. 4. Hardness data and polynomial fit of varying composition of MWCNT in PC. The cubic polynomial fit parameters are tabulated in Table 2.

Polynomial expression which is used to give a least square fit to measured data for hardness number using nano-indenter with concentration in % $x$ of MWCNT is given

$$H = \alpha + \beta x + \gamma x^2 + \delta x^3 \qquad (4)$$

Table 2. Coefficients , $\beta$ , $\gamma$ , $\delta$ (Eq.4)

| $\alpha$ | $\beta$ | $\gamma$ | $\delta$ |
|---|---|---|---|
| 91.47281 | 78.4298 | -5.60431 | 0.06775 |

We also present the polynomial fits to elastic modulus as well as maximum penetration depth caused due to nano- indentation in Figs. 5 and 6.

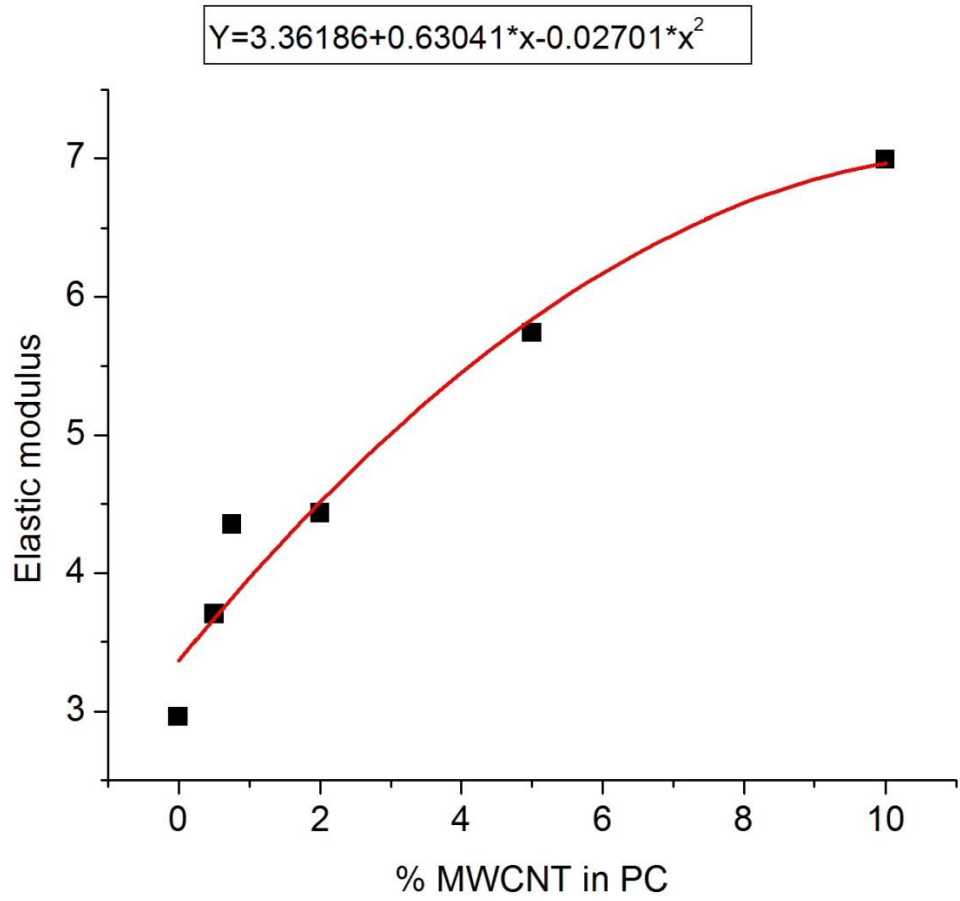

Fig.5 Elastic modulus and polynomial fit of varying composition of MWCNT in PC.

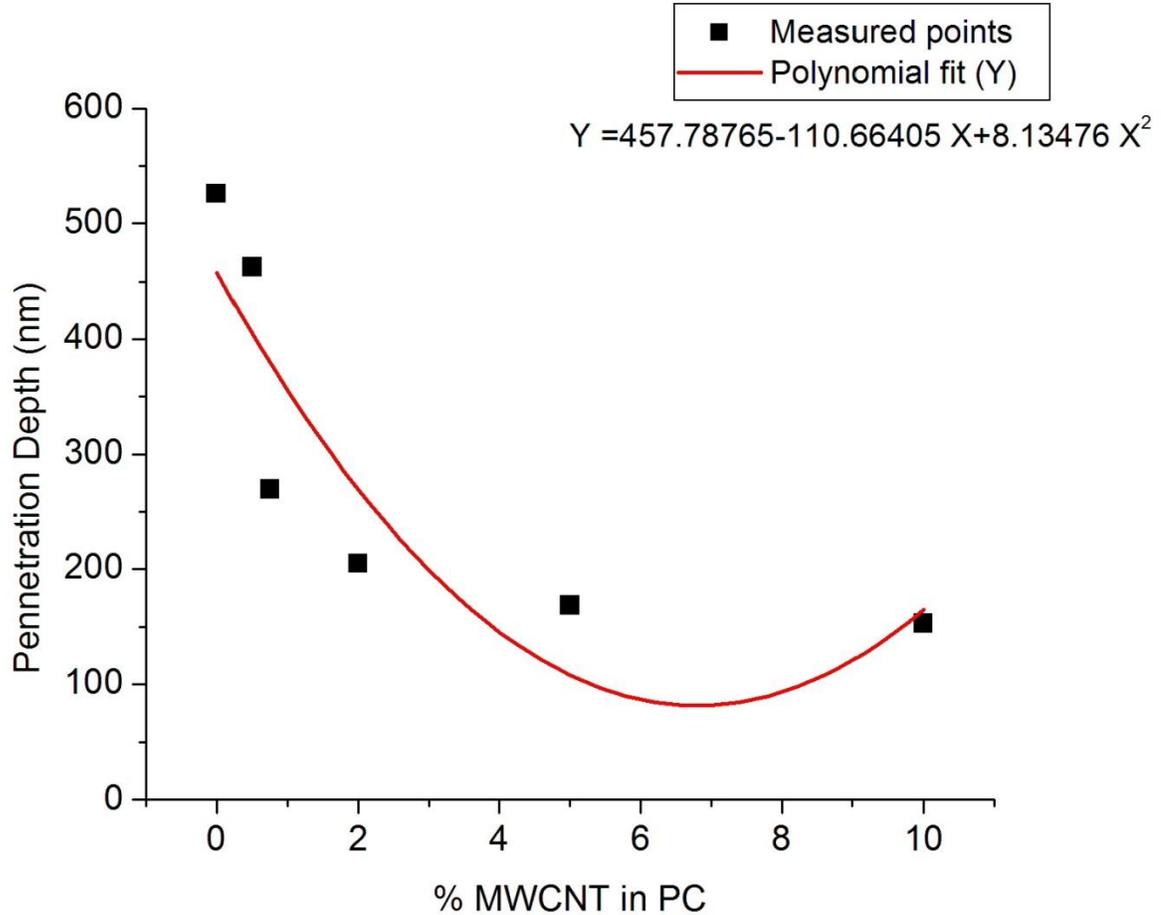

Fig.6 Penetration depth and polynomial fit of varying composition of MWCNT in PC.

### 4. Theoretical model for polycarbonate

Once we have presented the dynamical and static strength results as obtained from SHPB and nano-indenter, we now describe a simple model for polycarbonates based on Lenard-Jones (LJ) potentials. This model is a preliminary first step to understand interactions in PC and its composite. Polycarbonate is made up of Carbon(C), Hydrogen(H) and Oxygen(O) with molecule bisphenol-A as represented by $C_{16}H_{14}O_3$ and structure shown in Fig. 7.

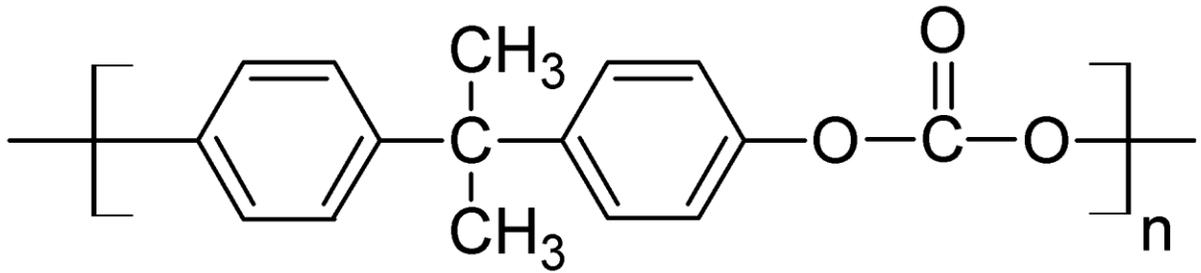

Fig.7 General structure of a PC molecule

Based upon various bond lengths which are usually around 0.142nm for C-C, nearly 10 bond lengths are involved in the length of a single unit of the molecule, making the length L as ~1.4nm. Similarly, the diameter D≈0.3nm. The total mass of the molecule $M_{pc}$~254.3 a.m.u. Some other important characteristics of PC material are given in Table 3.

Table 3. Some properties of polycarbonate

| | |
|---|---|
| ρ (density) | 1.2-1.22 gm/cc |
| Y (Young's modulus) | 2-2.4 GPa |
| $v_s$ (sound velocity) | 2270 m/s |
| $T_m$ (melting temperature) | 155-225°C |
| α (Thermal expansion Coeff.) | 65-70x10$^{-6}$/K |
| $C_v$ (Heat Capacity) | 1.2-1.3 KJ/Kg.K |
| κ (Thermal Conductivity) | 0.19-0.22 W/mK |

**4.1 Preliminaries**

Based upon the properties of PC material, we estimate the average inter-molecular distance $r$ in bulk PC as follows:

If $N_{pc}$ is the number of PC molecules in a volume $V$, then average volume associated with a single molecule of effective radial separation r is

$$\frac{4\pi r^3}{3} = \frac{V}{N_{pc}} \tag{5}$$

Effective volume of a single PC molecule can also be written in terms of bulk density ρ as $M_{pc}/\rho$. Thus

$$r = \left(\frac{3}{4\pi} M_{pc}/\rho\right)^{1/3} \tag{6}$$

The value using the characteristics as given in Table 3 result in $r$ ~0.44nm. Thus average inter-molecular distance is $2r$ ~0.88nm, which is about 0.63 of the length L. This estimate helps us in fixing the limits of variation of r in formulating our model as described below.

### 4.2. Interaction Energy

The molecules in a material interact through the atoms of the molecules. Generally these inter-molecular interactions are very weak and are governed by Van-der-Waal potentials. Following Pertsin et al[21], the molecule-molecule potential is based upon a summation over atom-atom potentials. Considering two molecules $Z_1$ and $Z_2$, one centered at the origin and oriented along z-axis, the second molecule situated at a position vector Rl and oriented with respect to origin atom and orientation expressed in terms of spherical polar coordinates as $(Z_2, \theta, \varphi)$. As shown in Fig. 8, we build up the atom-atom distances involved in the interaction.

Assuming molecules of length L and diameter D with $L \gg D$, we write interaction energy between any two molecules of PC (Fig. 8).

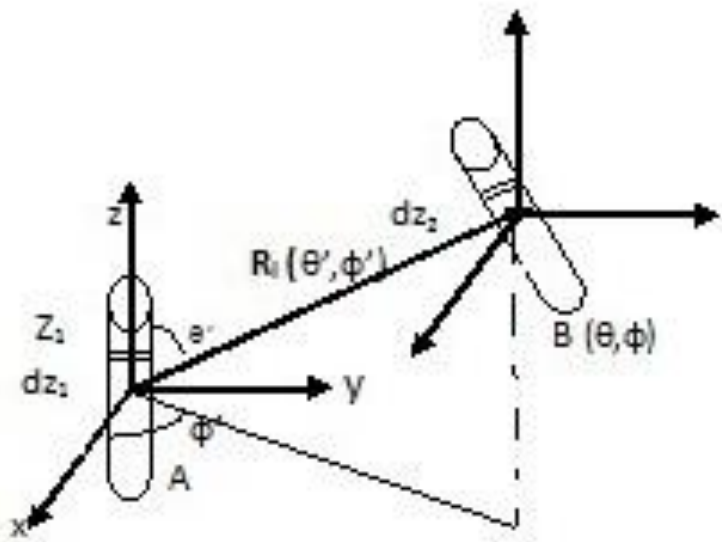

Fig.8  Molecules A and B and their coordinates.

The distance between two segments of molecules at $Z_1$ and $Z_2$ of lengths $dZ_1$ and $dZ_2$ is given by

$$r(z_1, z_2) = |R_l + z_2 - z_1| \tag{7}$$

The incremental lengths $dz_1$ and $dz_2$ have some fraction of C, H and O atoms assuming these are uniformly distributed over the total length L.

The distances R and $z_2$ are required to be expressed in Cartesian coordinates through their respective orientations as primed and unprimed angles,

$$R_x = R \sin\theta' \cos\varphi' \tag{8}$$

$$R_y = R \sin \theta' \sin \varphi' \tag{9}$$

$$R_z = R \cos \theta' \tag{10}$$

And for $z_2$ as

$$z_x = z_2 \sin \theta \cos \varphi \tag{11}$$

$$z_y = z_2 \sin \theta \sin \varphi \tag{12}$$

$$z_z = z_2 \cos \theta \tag{13}$$

$Z_1$ has been already taken along its z-axis. This helps in obtaining r, the distance between two sections of the interacting molecules by noting that eventually r is given by the total Cartesian as obtained in Eq.(6)

$$r^2 = x^2 + y^2 + z^2 \tag{14}$$

Assuming LJ potential between any two atoms I and j separated by a distance $r_{ij}$ as

$$u_{ij} = -\frac{A_{ij}}{r_{ij}^6} + \frac{B_{ij}}{r_{ij}^{12}} \tag{15}$$

where $A_{ij}$ and $B_{ij}$ are potential parameters which depend on the types of atoms involved in the interaction. For our molecule, the interactions involve C, H and O atoms. The relevant potential parameters obtained on the basis of a large number of aromatic hydrocarbons data are presented in Table 4.

Table 4. LJ potential parameters as obtained by self consistent fitting to aromatic hydrocarbons[22].

| Self-consistent Lennard-Jones 6-12 parameters | A / kcal mol-1Å$^6$ | B / kcal mol-1Å$^{12}$ |
|---|---|---|
| C-C | 1228.800000 | 2516582.400 |
| C-O | 754.059521 | 820711.722 |
| C-H | 79.857949 | 29108.222 |
| O-O | 429.496730 | 230584.301 |
| H-H | 2.56 | 81.920 |
| O-H | 39.075098 | 6035.457 |
| H-C | 79.857949 | 29108.222 |

Therefore, based on atom-atom interaction as defined in Eq. (7), the molecule-molecule potential energy between two sections of lengths $dz_1$ and $dz_2$ of an origin molecule and another at $R_l$ in an orientation of $z_2$ given by a polar angle θ and an azimuth angle φ is given by

$$U_{dz1,dz2} = \sum_{C,H,O} u_{ij} \tag{16}$$

where summation over i and j is limited to number of atoms in length dz of both molecules. Assuming the distance between these incremental sections of the molecules is r as in Eq. 6, this summation results in defining new potential parameters A and B as follows:

$$\langle A_T \rangle = 16^2 A_{CC} + 14^2 A_{HH} + 3^2 A_{OO}\ 2(16x14)A_{CH} + 2(16x3)A_{CO} + 2(14x3)A_{HO} \tag{17}$$

With a similar definition for $\langle B_T \rangle$. The number 16, 14 and 3 are the number of C, H and O atoms respectively in the whole molecule of PC. Based on these new parameters, it is now easy to express the total interaction energy between the two molecules in consideration as

$$U(R_l(\theta, \varphi)) = -\langle A_T \rangle \frac{1}{L^2} \iint_{-L/2}^{L/2} \frac{dz_1 dz_2}{r_{z_1 z_2}^6} + \langle B_T \rangle \frac{1}{L^2} \iint_{-L/2}^{L/2} \frac{dz_1 dz_2}{r_{z_1 z_2}^{12}} \qquad (18)$$

Measuring all distances in units of length L, the integrations become dimensionless numbers. Where the double integration is carried over the lengths of the two interacting molecules with number per unit length of each type of molecules has entered based on the fraction dz/L. The orientation angle of the 2nd molecule with respect to the origin molecule enters through the distance $r_{z1z2}$.

We carry out these integrations numerically at various centre orientations and distance of separation, R.

### 4.3 Numerical Results for pure PC

We have carried out numerical calculations for the interaction energy as detailed in the previous section. For numerical evaluation of the double integral (Eq. (10)), it is convenient to rewrite it in terms of dimensionless distances. If the two interacting molecules at the sites A and B are different, e.g. in case of MWCNT-PC interaction, we would use them of different lengths $L_A$ and $L_B$. Eq. (10) then reduces to

$$U(R_l(\theta, \varphi)) = -\langle A_T' \rangle / L_A^6 \iint_{-\frac{1}{2}}^{\frac{1}{2}} \frac{\frac{dz_1}{L_A} \frac{dz_2}{L_B}}{\frac{r_{z_1 z_2}^6}{L_A^6}} + \langle B_T' \rangle / L_A^6 \iint_{-\frac{1}{2}}^{\frac{1}{2}} \frac{\frac{dz_1}{L_A} \frac{dz_2}{L_B}}{\frac{r_{z_1 z_2}^{12}}{L_A^{12}}} \qquad (19)$$

Where now

$$r(z_1, z_2) = \left| R_l + \frac{L_B}{L_A} z_2 - z_1 \right| \qquad (20)$$

And the interaction parameters $\langle A'_T \rangle$ and $\langle B'_T \rangle$ are same as defined in Eq.(9) if interacting molecules are PC, but are defined as below in case the interacting molecules are PC and MWCNT,

$$\langle A'_T \rangle = N_{mwcnt}(16 A_{CC} + 14 A_{HH} + 3 A_{OO} + 14 A_{CH} + 3 A_{CO}) \tag{21}$$

With similar definition for $\langle B'_T \rangle$. Here $N_{mwcnt}$ is the number of C atoms in one MWCNT.

In what follows the length of PC molecule is defined as L instead of $L_A$.

The numerical integrations are carried out using Simpson's rule and by varying R, the distance between centres of two interacting molecules as well as by varying angles of spherical polar coordinates to determine orientations. As an example, an angle θ =0, which is polar angle indicates parallel molecules whereas θ =π/2 indicates perpendicular orientation. A random orientation would be close to θ =π/4. We have chosen two different orientations, one for determining the position of the centre of the B molecule with respect to A and the other for the orientation of B molecule in its own centre coordinates.

A typical representative molecule-molecule interaction energy at various separation distance R (expressed in PC molecule length L=1.4nm) thus obtained at some orientation of R and B molecule is shown in Fig. 9, 10, 11. These representative figures clearly indicates that the minimum energy at each orientation will determine the inter molecule separation and thus the volume and density of the solid material, which can be compared with the measured density.

It is important to emphasize that in obtaining the numerical results, we have used the available potential parameters to determine PC-PC interaction energy and other interaction energies for PC-MWCNT. No adjustment or fitting to these parameters has been done. The magnitude of energy determines the melting temperature and its second derivative with respect to distance at minimum energy determines the elastic constant.

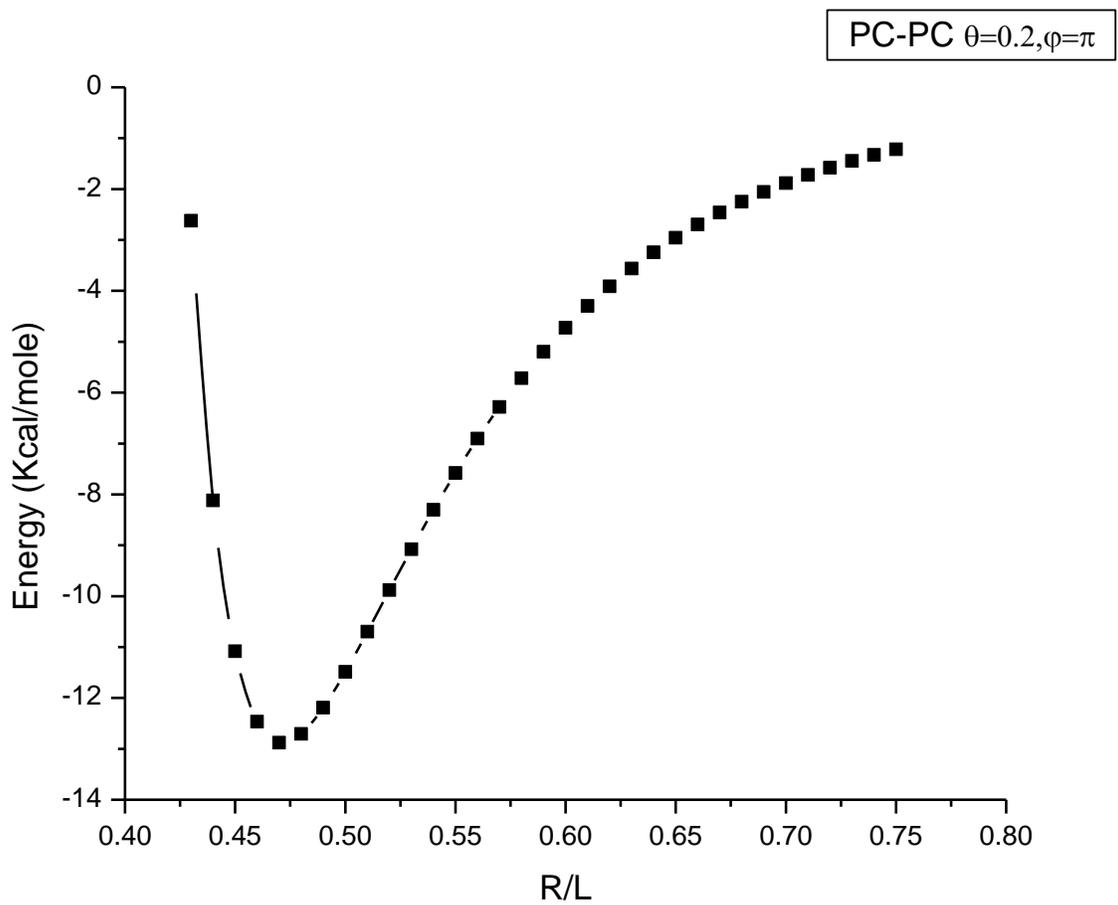

Fig.9 PC-PC molecule interaction energy calculated by using LJ potential as a function of the distance of separation at some orientation as mentioned.

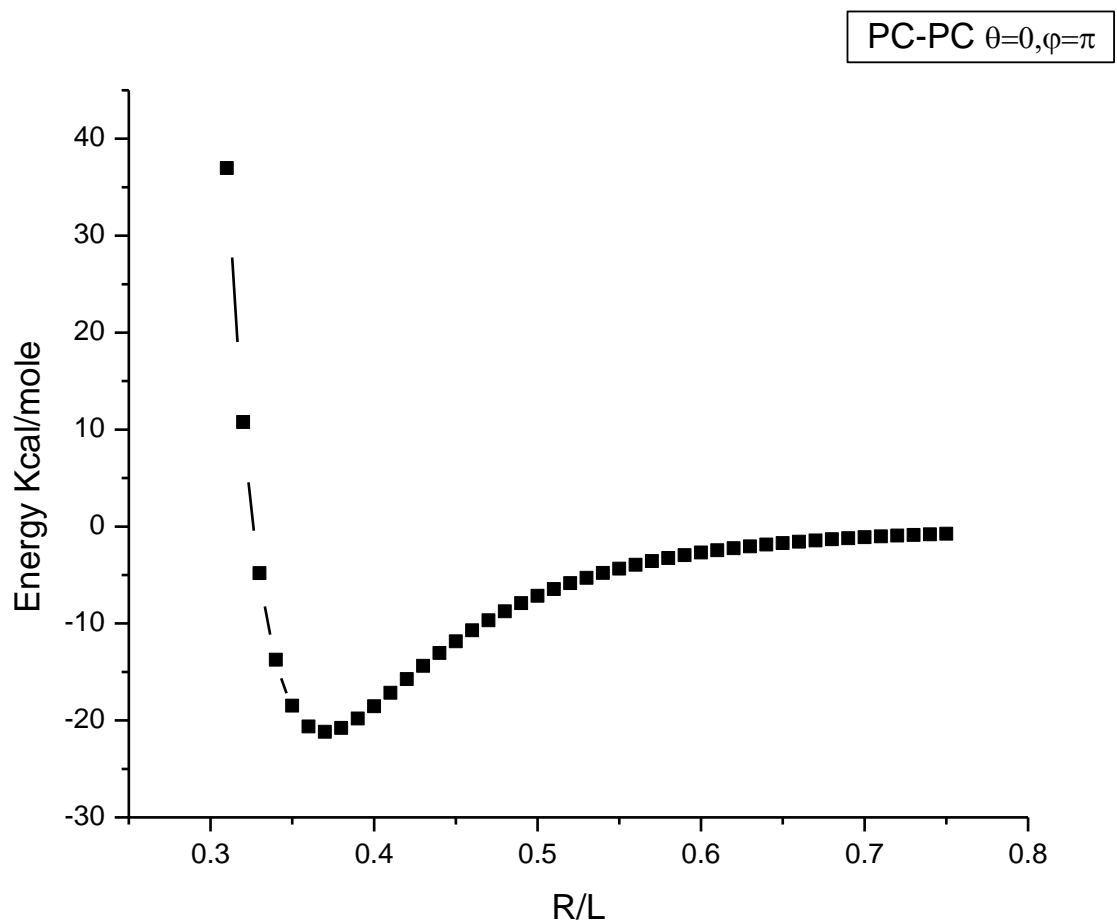

Fig.10 PC-PC molecule interaction energy calculated by using LJ potential as a function of the distance of separation at some orientation as mentioned.

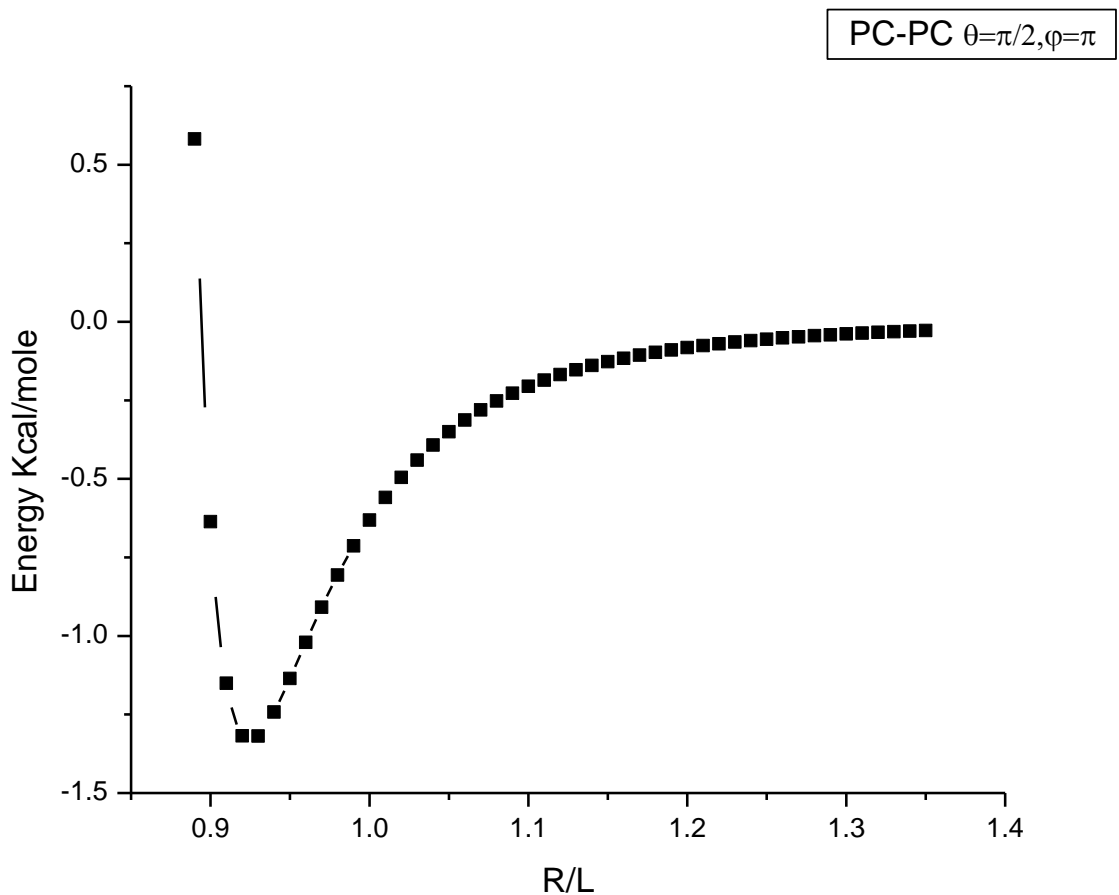

Fig. 11 PC-PC molecule interaction energy calculated by using LJ potential as a function of the distance of separation at some orientation as mentioned.

Similarly, representative plots (Fig.12,13) of calculated density can be obtained from the minimum energy values are shown at two different orientations.

Finally, we present in Fig.14 a combined plot of PC-PC interaction energy at 3 different orientations to have an estimate of various calculated parameters like minimum energy, separation distance, density and the second derivative of the energy which is related to the Young's modulus.

We tabulate the computed values of computed structural characteristics using our model in Table-5.

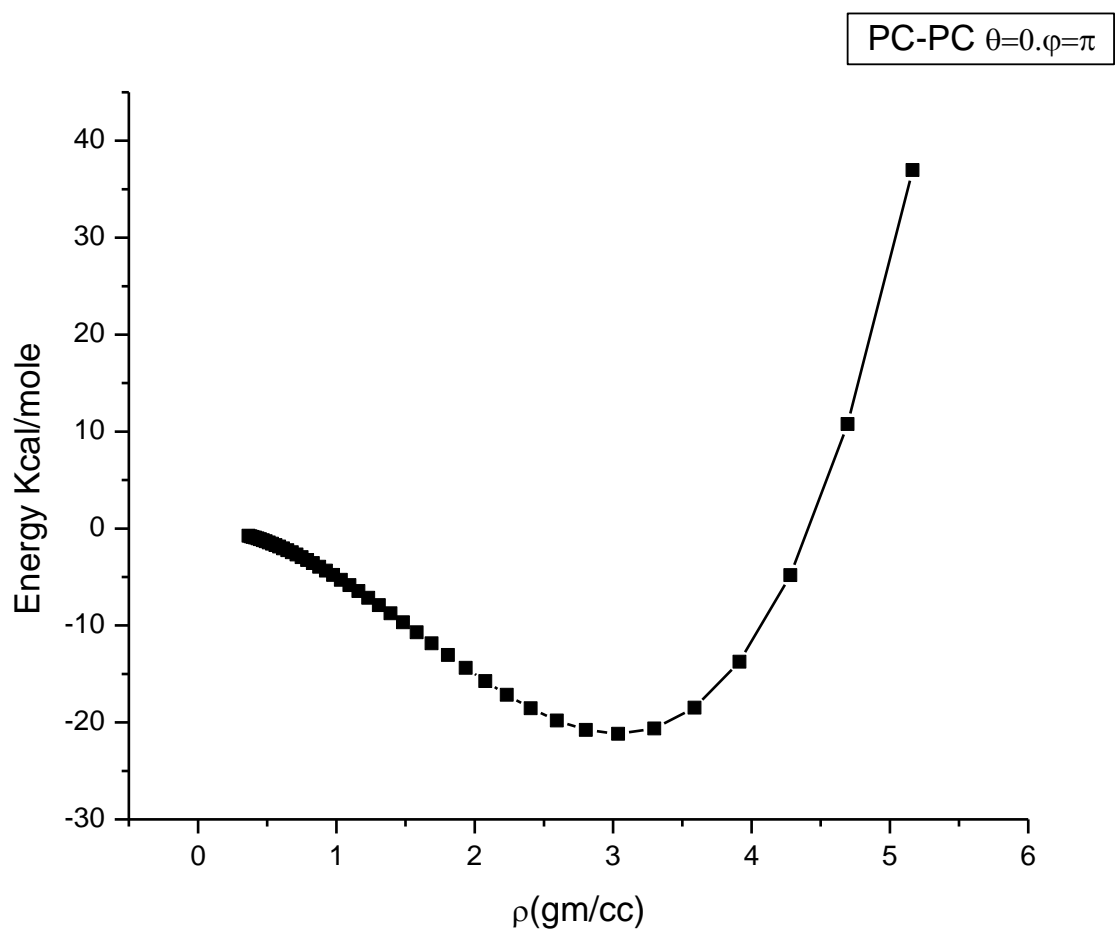

Fig. 12 PC-PC molecule interaction energy calculated by using LJ potential as a function density for the mentioned orientation

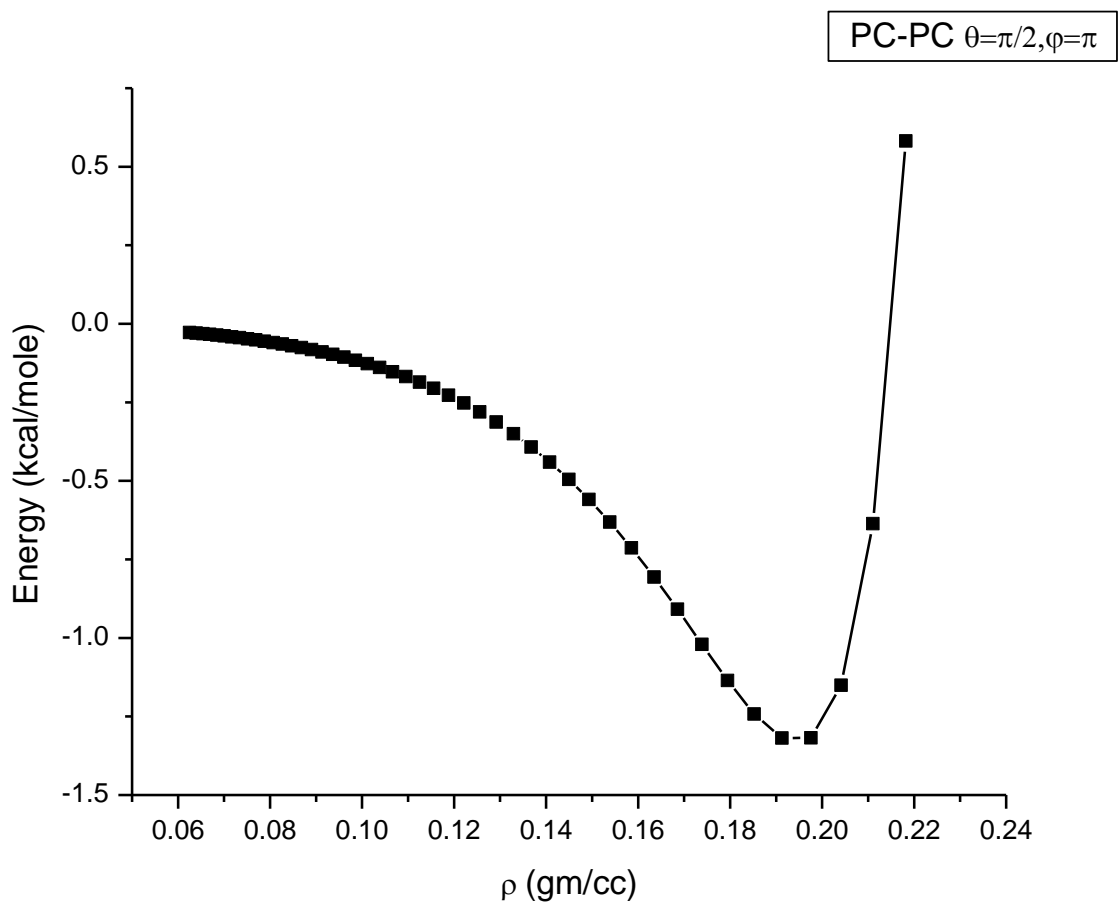

Fig. 13 PC-PC molecule interaction energy calculated by using LJ potential as a function density for the mentioned orientation

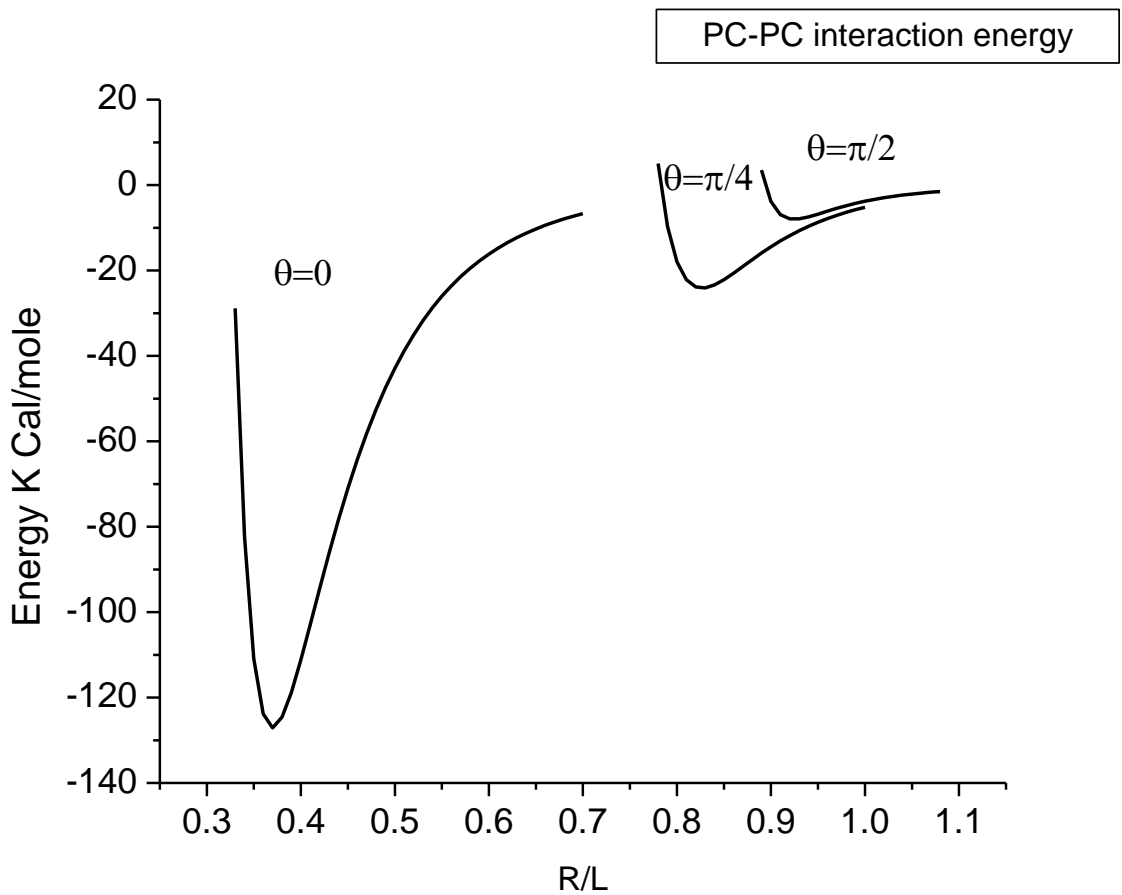

Fig.14 Variation of PC-PC interaction energy as a function of distance and orientation.

Table 5. Computed characteristics of pure PC.

| Polar angle $\theta$ units of $\pi$ rad | R/L | $\rho$ (gm/cc) | E (Kcal/mole) | Y (GPa) |
|---|---|---|---|---|
| 0 | 0.38 | 11.2 | -124.6 | 16.8 |
| 1/4 | 0.83 | 1.06 | -23.6 | 1.43 |
| 1/2 | 0.93 | 0.76 | -7.9 | 0.51 |

We notice that at orientations around π/4, the computed results for density, energy and Y are reasonably close to the observed values as given in Table 3. In fact, one can easily understand the effect of orientation, when the molecules are parallel (θ =0), the molecules bind very strongly as there is interaction between the whole length of one molecule with the whole length of the other molecule, whereas for perpendicular orientation (θ = π/2), interaction is weakest and results in low density and small value of Y. The optimum orientation in randomly ordered PC solid is when an intermediate orientation is chosen.

5. **Model and numerical results for MWCNT-PC composite**

A composite of PC can be modeled on the basis of a mixture of appropriate number of PC-MWCNT and PC-PC interaction energies where the number of MWCNTs is chosen in accordance with the weight percentage of MWCNTs forming the PC composite. As a demonstrative example we choose 5% composition and establish our model.

Using similar interaction potentials, except that the atoms of PC which are 16 of C type, 14 of H type and 3 of O type interact with all C atoms of MWCNT.

An MWCNT of diameter d and length $L_B$ has large number of C atoms and is determined by finding out the number of hexagons in its area. Further, there are several walls of an outer diameter d tube, successively added on to lower diameters which continue to decrease by 2x0.34nm, 0.34 nm being the intertube separation.

A single wall will have twice the number of hexagonal areas as number of C atoms. This will therefore be

$$N_{cnt} = \frac{2\pi dL}{\frac{3\sqrt{3}b^2}{2}} \qquad (22)$$

Where b is the C-C bond length of 0.142nm. The number of total carbon atoms in MWCNT will be obtained when the number of walls is multiplied with average diameter.

Since our MWCNTs were of an average length of 60 nm and also of diameter of 60 nm, this number is easily calculated.

The number of a single wall of our dimensions of MWCNT comes out to be around 431770.

Assuming 10 walls, the mass of a single MWCNT is around $8.6 \times 10^{-17}$ gm in comparison to that of a molecule of PC of $4.2 \times 10^{-20}$ gm. This allows us to choose the ratio of MWCNT number for some PC molecules to account for required weight % of MWCNT composition. In this way, we combined the interaction energy by appropriate weight factor for a correct %.

We present our results of the composite thus obtained in the Figs. 15, 16, 17.

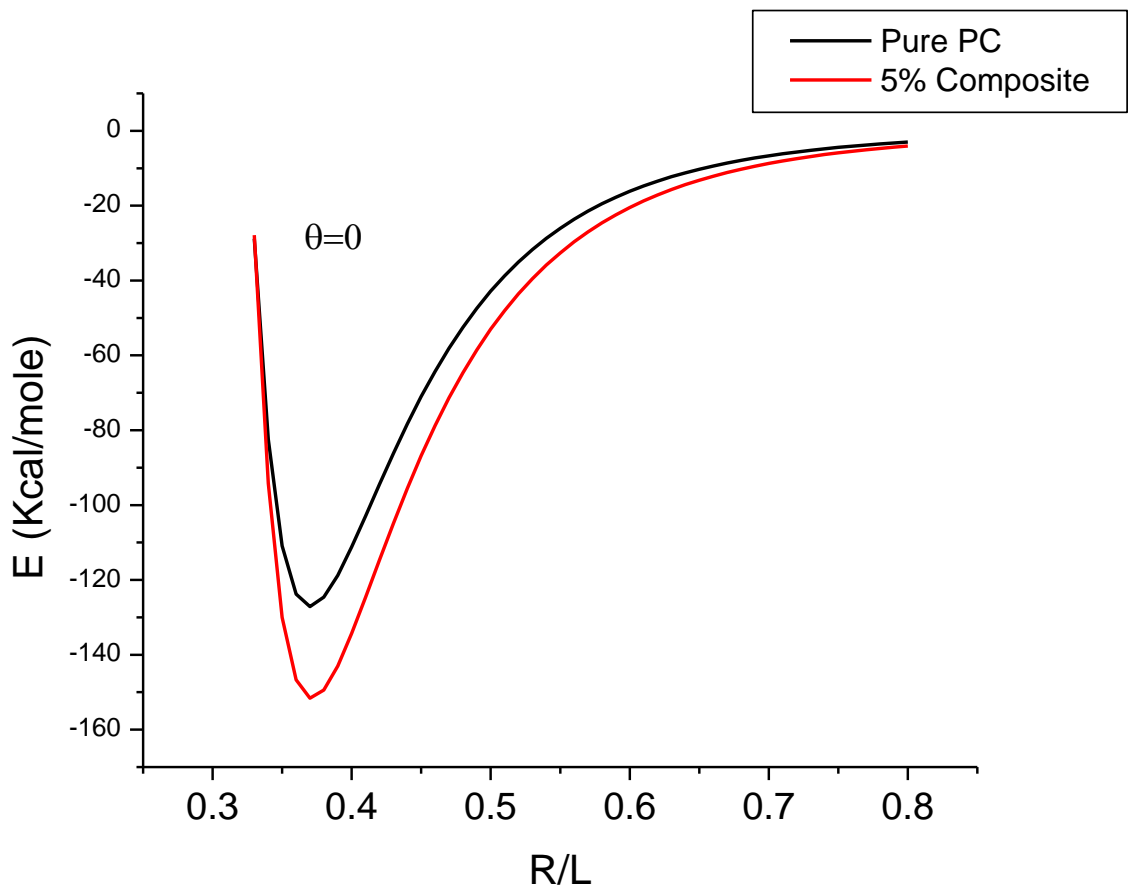

Fig. 15 Variation of PC combined with 5% by weight of MWCNT interaction energy as a function of distance and orientation parallel.

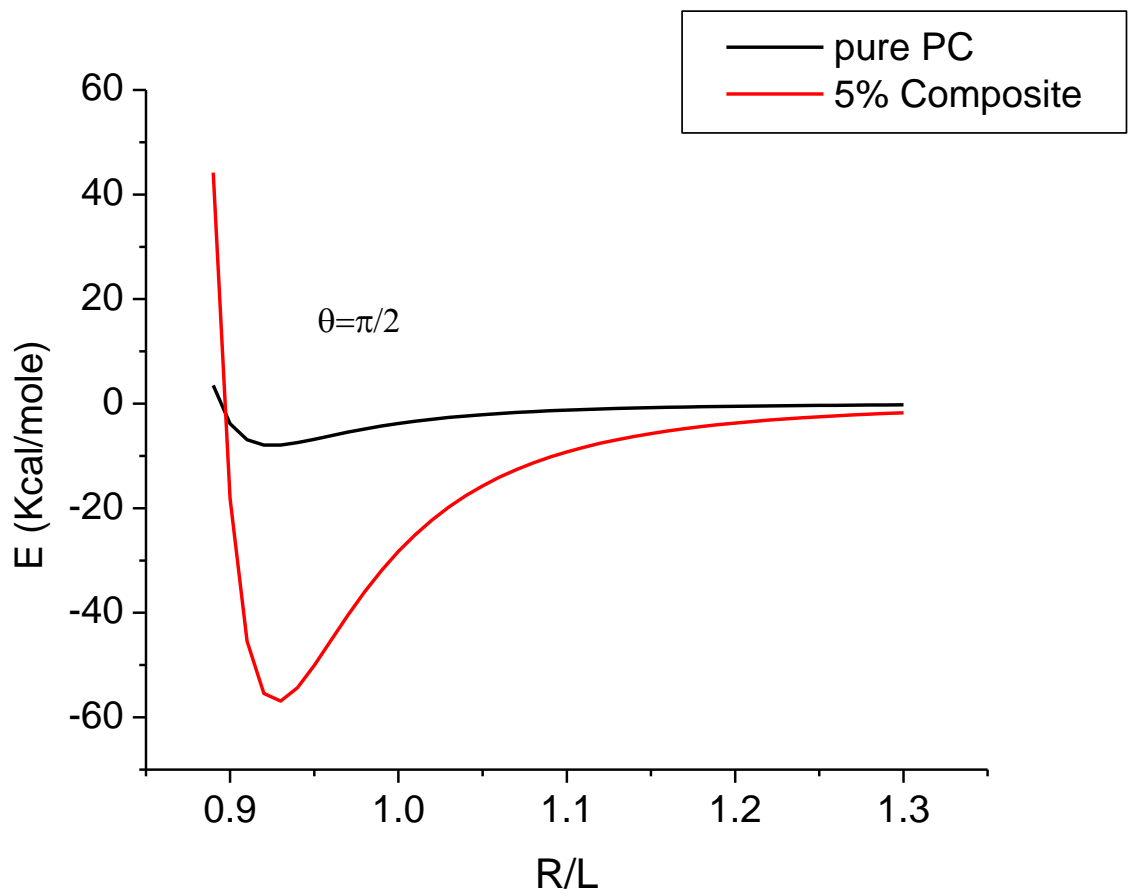

Fig.16 Variation of PC combined with 5% by weight of MWCNT interaction energy as a function of distance and orientation perpendicular to each other.

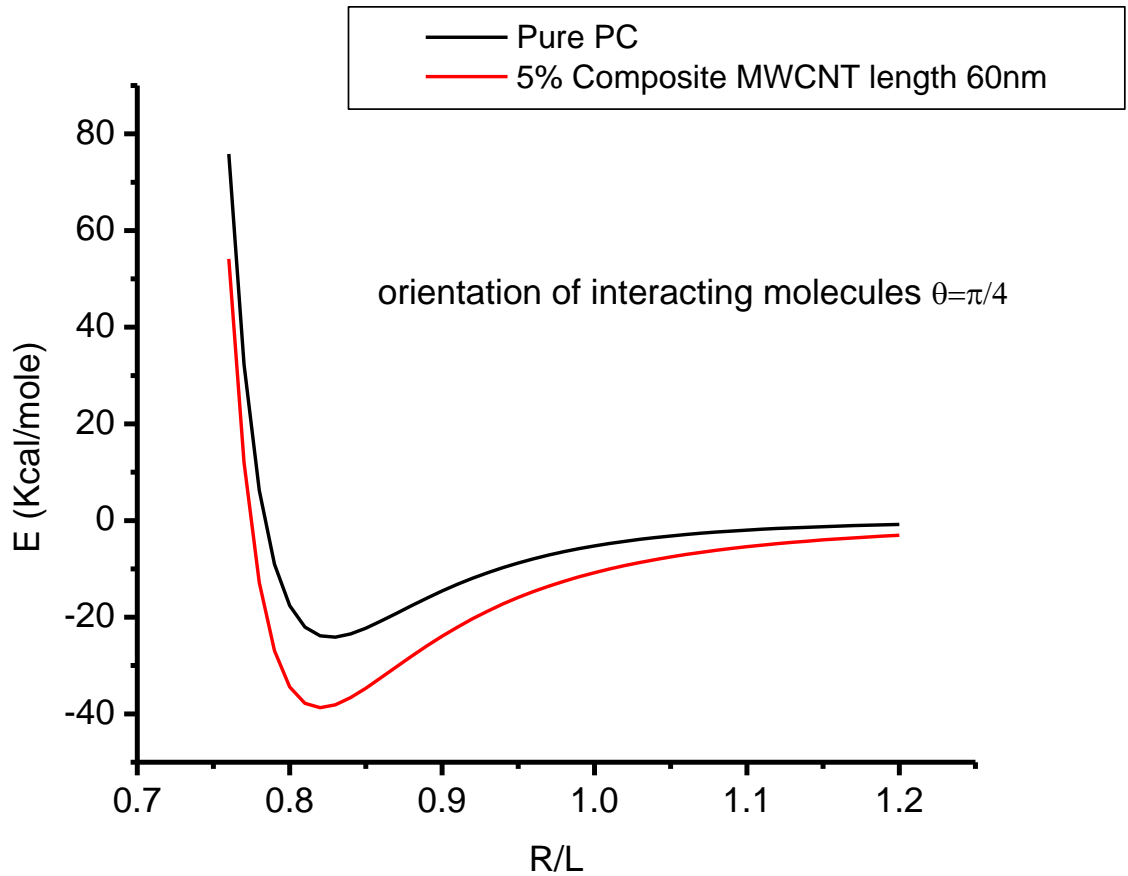

Fig.17 Variation of PC combined with 5% by weight of MWCNT interaction energy as a function of distance and orientation chosen as average of parallel and perpendicular between each other.

The results thus obtained for θ=π/4 for PC-MWCNT (5%) are presented in Table.6

Table . PC-MWCNT (5%) composite model characteristics

| Polar angle θ units of π rad | R/L | ρ (gm/cc) | E (Kcal/mole) | Y (GPa) |
|---|---|---|---|---|
| 1/4 | 0.82 | 1.12 | -38.7 | 3.36 |

It is easily observed from this Table that there is significant enhancement of about 60% in cohesive energy and the Young's modulus increases almost by a factor of 2.

## 6. Discussion and concluding remarks

This work presents two kinds of results. Firstly all the measured experimental results on stress-strain related characteristics using either dynamical impact or static impact have been presented in equations which are representative of all these results. The equations presented here summarize the effect of composition of MWCNTs on dynamic impact resistance, even realizing what compositions yield optimization for maximum benefit in favor of enhancement of the PC by forming composites. The results when fitted to static characteristics yield different composition for maximizing the enhancement. Indeed, it is found that higher compositions even above 5% are more important for static properties in comparison to about 2% limit for the case of dynamic impact.

More importantly, we present a simple model to represent MWCNT based composites. It is very satisfying to notice that on the basis of an assumption of a simple model potential between constituent atoms of the material whose form is taken as 6-12 LJ potential and whose interacting parameters are picked up from literature and not tailored to the chosen material, is able to qualitatively interpret the properties of bulk PC and its composites. We do not attempt any adjustment of the potential or structure as our aim has been to understand in a simple way if the strength of the composite material gets modified in the manner the experimental observations are made. In that goal, we have succeeded. We need to modify our procedure by incorporating kinetic pressure terms in the interaction to make them suitable for dynamic impact study. However, as a first step this calculation and model provides enough insight into the problem. Therefore the proposed model and procedure can be refined to optimize the observed static and dynamic results for various mechanical properties of PC-MWCNT composites of various compositions.